\documentclass[11pt]{article}

\usepackage[final]{acl}

\usepackage{times}
\usepackage{latexsym}
\usepackage{amsmath}
\usepackage[T1]{fontenc}
\usepackage{booktabs}

\usepackage[utf8]{inputenc}

\usepackage{microtype}

\usepackage{inconsolata}

\usepackage{graphicx}

\title{Beyond Text-to-SQL: Autonomous Research-Driven Database Exploration with DAR}

         \author{
  Ostap Vykhopen,
  Viktoria Skorik,
 Maksym Tereshchenko \and
  Veronika Solopova \\
  \textbf{Correspondence:} \href{mailto:ostap.vykhopen@mantisanalytics.com}{ostap.vykhopen@mantisanalytics.com}
}

\begin{document}
\maketitle
\begin{abstract}
Large language models can already query databases, yet most existing systems remain reactive: they rely on explicit user prompts and do not actively explore data. We introduce DAR (Data Agnostic Researcher), a multi-agent system that performs end-to-end database research without human-initiated queries. DAR orchestrates specialized AI agents across three layers: initialization (intent inference and metadata extraction), execution (SQL and AI-based query synthesis with iterative validation), and synthesis (report generation with built-in quality control). All reasoning is executed directly inside BigQuery using native generative AI functions, eliminating data movement and preserving data governance. On a realistic asset–incident dataset, DAR completes the full analytical task in 16 minutes, compared to 8.5 hours for a professional analyst (approximately $32\times$ faster), while producing useful pattern-based insights and evidence-grounded recommendations. Although human experts continue to offer deeper contextual interpretation, DAR excels at rapid exploratory analysis. Overall, this work shifts database interaction from query-driven assistance toward autonomous, research-driven exploration within cloud data warehouses.
\end{abstract}

\section{Introduction}
Modern enterprises accumulate data at unprecedented scales, yet much of it becomes ``dark data'', poorly indexed, rarely queried, and eventually lost~\cite{schembera2019dark,heidorn2008shedding}. Industry estimates suggest that over 90\% of organizational data remains unused, representing a major missed opportunity~\cite{george2023bringing}. The bottleneck is no longer computation: modern warehouses answer complex queries in seconds, but require humans to notice opportunities, prepare data, and manually craft queries~\cite{kuchnik2022plumber}. As data collections grow, this effort scales poorly, leaving patterns undiscovered and cross-table relationships unexplored. 

What enterprises need is autonomous, proactive exploration rather than purely reactive querying. AI-powered data exploration promises a shift from responding to isolated questions to running continuous research cycles that identify promising directions and follow-up analyses~\cite{zheng2025automation}.

To address these challenges, we present Data Agnostic Researcher (DAR), a multi-agent system for autonomous exploration of relational databases using BigQuery’s native generative AI functions. DAR is instantiated on Google Cloud with Gemini-based agents that plan research goals, synthesize and execute SQL+AI queries, and iteratively generate reports entirely inside the warehouse. We study three research questions:
\begin{enumerate}
    \item RQ1: How can a hierarchical multi-agent architecture support end-to-end autonomous database research without manually specified queries? 
    \item RQ2: To what extent can native in-database LLM functions replace external LLM calls while preserving data security and latency? 
    \item RQ3: How does an autonomous system like DAR compare with professional human analysts in analysis time and usefulness of insights on realistic business-intelligence tasks?  
\end{enumerate}

 Our main contributions are threefold: (i) we introduce DAR, a multi-agent architecture for proactive database exploration that formulates its own research questions, generates and validates SQL+AI queries, and discovers patterns without human-initiated prompts; (ii) we provide a fully in-database implementation on BigQuery using only native generative AI functions, covering the entire pipeline from schema discovery to report generation with quality-control loops and releasing it as open source; and (iii) we empirically compare DAR with a professional human analyst on an asset--incident dataset, showing an approximate $32\times$ reduction in total analysis and reporting time while highlighting the complementary role of in-depth human interpretation.
\section{Related work}
\subsection{Multi-Agent Systems and LLM Agents}
Recent advances in LLM-based agents have demonstrated capability to perceive environments, make autonomous decisions, and orchestrate tools to accomplish complex tasks \cite{zhou2025autonomous}. Multi-agent systems enhance the abilities of single agents by coordinating and collaborating, allowing agents to communicate and share problem-solving tasks across specialized roles.

SchemaAgent pioneered multi-agent frameworks for database schema generation, demonstrating how specialized agents can collaborate on database-related tasks through role assignment and error detection mechanisms \cite{wang2025text2schema}. However, this system focuses on schema design rather than autonomous data exploration. 

Beyond schema-focused agent frameworks, prior work on multi-agent decision-making typically assumes explicit task definitions and engineered reward structures. However, traditional multi-agent reinforcement learning methods need a lot of training and specific reward frameworks \cite{zhang2019multiagent}, while autonomous exploration demands systems that proactively investigate without explicit reward signals.

\subsection{Database Querying and Understanding}
Text-to-SQL systems translate natural language queries into SQL commands, with modern LLM-based approaches demonstrating strong performance even on complex structures through retrieval-augmented generation, database metadata, and self-correction loops \cite{deng2022texttosql}. However, text-to-SQL systems remain fundamentally reactive. Designed to answer explicit user questions rather than proactively exploring databases to uncover patterns, relationships, or insights. They excel at query generation but cannot formulate their own research questions or decide what to investigate.

Beyond query generation, current database agents can list tables, retrieve schemas, and execute queries \cite{hong2024nextgen,mohammadjafari2025natural}. Yet their focus remains on structural understanding: examining metadata, tables, and columns. They rarely explore data content, quality characteristics, or statistical patterns across tables. Autonomous exploration requires investigating what the data contains, not just how it's organized.

Recent advances in cloud platforms like BigQuery have introduced native generative AI capabilities support functions like AI.GENERATE, AI.GENERATE\_BOOL, and AI.GENERATE\_TABLE for structured data extraction, along with row-wise LLM inference for SQL-driven data manipulation \cite{google2025bigqueryai}. BigQuery's evolution toward an autonomous data-to-AI platform includes automated metadata generation, Gemini-assisted data preparation, and multimodal ObjectRef data types \cite{google2025bigqueryautonomous}.

Existing agentic systems fail to utilize their inherent capabilities. Instead, they depend on external LLM API calls, which lead to issues with data transfer, delays, and security risks, rather than processing all reasoning within the database.
\subsection{Knowledge Graphs and Data Exploration}
GraphRAG methods show how LLM-generated knowledge graphs enhance retrieval-augmented generation through structured relationships and multi-hop reasoning \cite{hu2024grag}. Enterprise knowledge graphs built through LLM-based entity extraction support advanced querying across heterogeneous sources \cite{kumar2025llmkg}.

However, traditional knowledge graph approaches require explicit graph construction through data extraction and transformation, operating as separate graph stores rather than directly querying production relational databases. They require preprocessing steps like ETL pipelines and cannot dynamically investigate the original relational database without first materializing the data into the graph system.

\subsection{Research Gap}
Despite progress, no existing system combines the key capabilities needed for autonomous database exploration: (i) agentic workflows that proactively explore data without explicit queries; (ii) native in-database AI using warehouse generative functions to avoid data movement; (iii) multi-agent coordination for schema discovery, profiling, relationship mapping, and pattern mining; (iv) data-agnostic adaptation to arbitrary schemas; and (v) end-to-end workflows from initial exploration to insight generation.

Existing text-to-SQL systems remain reactive to user prompts~\cite{deng2022texttosql,hong2024nextgen}, schema discovery focuses on structure alone~\cite{marotta2018schema}, knowledge graphs are typically decoupled from databases~\cite{Mohamed2025KnowledgeGraphs,Tian2022KGReasoning}, and database agents largely wait for human instructions~\cite{Wang2024LLMAgentsSurvey}. To the best of our knowledge, no prior system autonomously formulates research questions, explores data content, and generates insights without human query initiation.

\section{Methods}
\begin{figure}[t]
    \centering
    \includegraphics[width=\columnwidth]{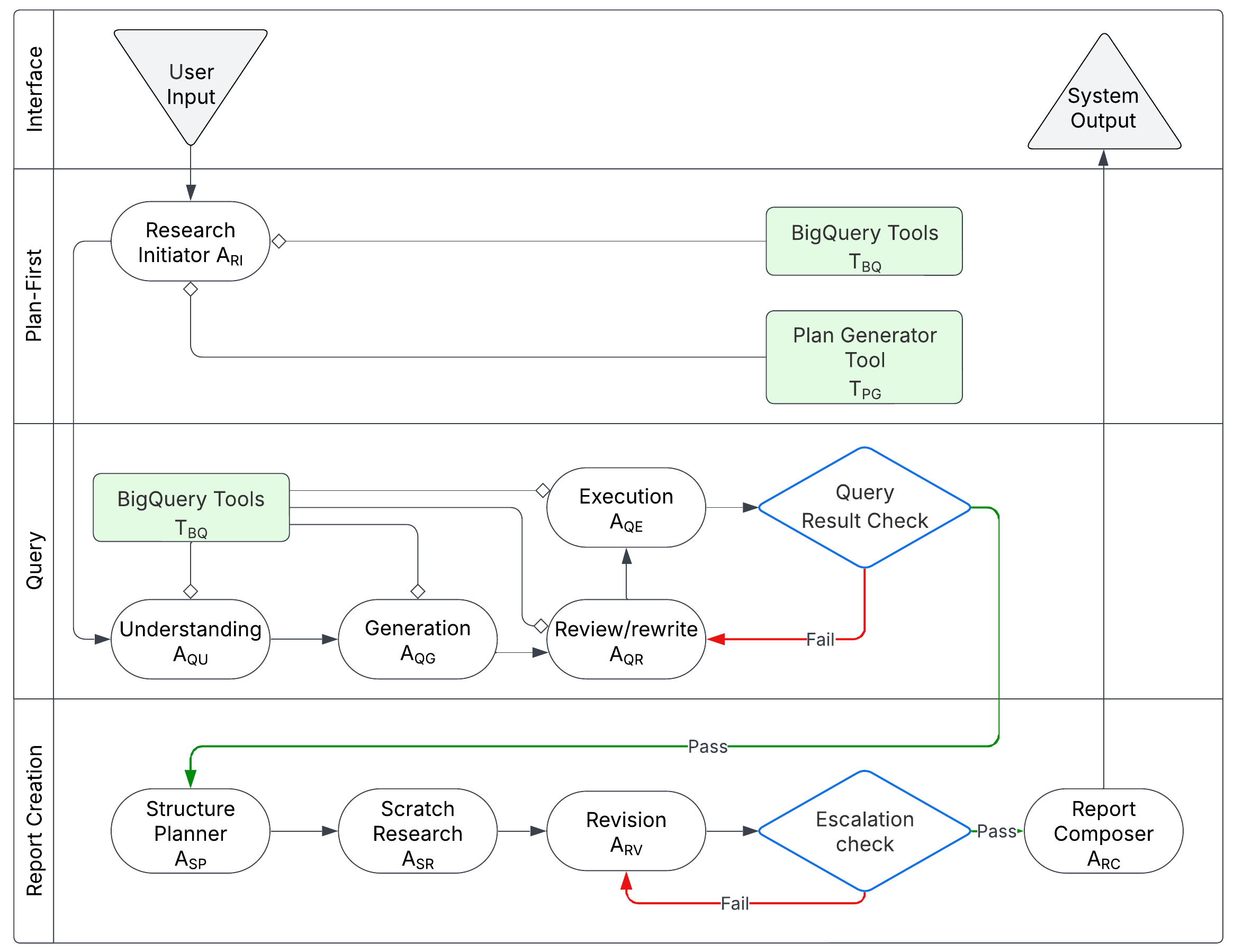}
    \caption{Performance comparison of the DAR system and human analyst.}
    \label{fig:DAR1}
\end{figure}

The proposed architecture embodies a hierarchical multi-agent system design where intelligence is distributed across specialized agents, each responsible for distinct aspects of the research pipeline. This modular approach ensures scalability, maintainability, and adaptability to diverse database contexts. The architecture is illustrated in Figure \ref{fig:DAR1}.
\subsection{System Architecture}

The DAR system adopts a three-layer hierarchical architecture designed to support end-to-end autonomous database exploration. At the Initialization Layer, the system processes user input and extracts database metadata, translating high-level research intents into a structured representation grounded in the available schema. The Execution Layer then operationalizes these intents by synthesizing, executing, and validating SQL queries, including those that invoke BigQuery’s native generative AI functions. Finally, the Synthesis Layer aggregates intermediate results and produces human-readable reports that summarize the discovered patterns and insights. These layers are connected through a sequential workflow orchestration mechanism augmented with conditional feedback loops, which enable error recovery, iterative refinement, and basic quality control without manual intervention.
\subsection{Agent Specification}
The overall architecture is instantiated as a set of specialized agents, each responsible for a well-defined subset of the research pipeline. In this subsection, we describe the main agent roles and their interactions.

 \textbf{Research Initiator Agent}. In our methodology, the Research Initiator agent ($A_{RI}$) acts as the primary orchestrator of the research process. It is responsible for parsing natural language problem statements, maintaining the workflow state across multi-turn sessions, and delegating tasks to specialized sub-agents through ADK’s AgentTool interface. Based on the user’s objectives and the available database metadata, $A_{RI}$ decomposes the overall research goal into executable subtasks and coordinates their execution.
 
The agent relies on two key auxiliary components. First, the meta extractor toolset provides schema intelligence via four metadata operations: \textit{list\_table\_ids} (enumerating tables), \textit{list\_dataset\_ids} (discovering datasets), \textit{list\_table\_info} (retrieving column schemas and constraints), and \textit{list\_dataset\_info} (accessing dataset-level metadata). Together, these tools give $A_{RI}$ a structured view of the database environment. Second, the plan generator tool ($T_{PG}$) synthesizes concrete execution strategies by jointly considering the research objectives, schema information, and resource constraints (e.g., query cost or time limits). The resulting plan specifies the sequence of SQL and analysis steps that downstream agents will carry out.

 \textbf{SQL Execution Pipeline}. The SQL execution pipeline is implemented as a four-agent sequential workflow with integrated feedback control. The Query Understanding agent ($A_{QU}$): first maps the high-level research objectives into queryable components, using the metadata supplied by the Meta Extractor to align user intents with specific tables, columns, and relationships. The Query Generation agent ($A_{QG}$) then constructs concrete SQL statements, optionally embedding BigQuery-native AI functions such as ML.GENERATE\_TEXT, AI.GENERATE, AI.GENERATE\_BOOL, AI.GENERATE\_TABLE, and AI.GENERATE\_DOUBLE to perform generative operations directly inside the database.
 
The resulting queries are handed to the Query Execution agent ($A_{QE}$), which manages the connection to BigQuery, submits the queries via the \textit{execute\_sql} tool, and retrieves the corresponding results using authenticated service account credentials. Finally, the Query Review agent ($A_{QR}$) analyzes execution failures or unsatisfactory results and, when necessary, generates revised queries. This agent implements an iterative refinement loop with an upper bound of iterations to prevent unbounded retry cycles. 

Query outcomes are passed through a validation stage that determines whether they are acceptable for downstream synthesis. A result is considered valid if it is non-empty and free of execution errors, i.e., the decision node. Result Validation he validator evaluates query outcomes as follows:
\[
\text{val} = \text{PASS}
\iff
\lvert \text{Result} \rvert > 0 \land \text{Error}(\text{Result}) = \emptyset.
\]
If validation fails, control is returned to $A_{QR}$ to trigger another refinement step; if validation passes, the pipeline advances to the report generation stage.

\textbf{Report Generation Pipeline}. The report generation pipeline is organized as a four-agent collaborative workflow with explicit quality gates. The Structure Planner  ($A_{SP}$) first defines the overall report architecture, including the section hierarchy and intended narrative flow, based on the research objectives and the validated query results. The Scratch Research agent ($A_{SR}$) then performs a first-pass synthesis of the findings, operating under strict information boundaries that restrict it to the outputs of the SQL execution pipeline and associated metadata. Subsequently, the Revision agent ($A_{RV}$) refines the draft report by addressing issues of coherence, readability, and completeness, guided by evaluation feedback. This refinement process is iterative but bounded by a maximum of j iterations to control computational cost. The final report is produced by the Report Composer ($A_{RC}$), which assembles the revised content into a deliverable artifact in Markdown format suitable for downstream rendering or export.
The decision node. The Escalation Checker routes outputs:
\[
\begin{cases}
\text{Quality}(\text{Report}) \ge \theta \;\Rightarrow\; \text{\small{Proceed to final formatting}},\\[4pt]
\text{Quality}(\text{Report}) < \theta \;\Rightarrow\; \text{\small{Return to Revision Agent}}.
\end{cases}
\]
An Escalation Checker serves as the decision node for this pipeline. It evaluates each report against an internal quality score and compares it to a threshold $\theta$. If the quality score meets or exceeds $\theta$, the report is accepted and forwarded to final formatting. If the score falls below $\theta$, the report is routed back to $A_{RV}$ for another refinement cycle. This mechanism provides a simple but effective quality control loop that does not require human review.

\subsection{Reasoning Engine: ADK-Based LLM Orchestration}
The reasoning and orchestration capabilities of DAR are built on top of Google’s Agent Development Kit (ADK), which serves as the foundational framework for defining, composing, and executing agents. This design choice offers several architectural advantages. First, model flexibility is preserved: ADK supports heterogeneous model backends, enabling DAR to primarily utilize Gemini models that are optimized for the Google Cloud ecosystem, while also allowing integration of alternative large language models via LiteLLM and, where available, fine-tuned domain-specific models. This flexibility is crucial for adapting DAR to different deployment environments and application domains without redesigning the agent logic. Second, the architecture leverages ADK’s agent composition patterns by combining three main agent types. LLMAgents are used for tasks that require open-ended reasoning, natural language understanding, and dynamic decision-making, such as the Research Initiator, Query Understanding, and report generation agents. Workflow agents implement structured process orchestration using SequentialAgent patterns, which allows deterministic control flows for the SQL execution and report creation pipelines. Custom agents encapsulate specialized capabilities and external integrations, including BigQuery connectors and metadata extractors, thereby separating infrastructure concerns from high-level reasoning. Third, DAR incorporates multiple reasoning strategies to improve robustness and interpretability. Several agents employ chain-of-thought (CoT) prompting to explicitly articulate intermediate reasoning steps when dealing with complex queries or planning multi-step workflows. A ReAct-style (Reasoning + Acting) pattern is used to interleave deliberation with tool invocation, allowing agents to reason about which tools to call, interpret the results, and revise their strategy in a closed loop. Finally, self-reflection is built into components such as the Query Review and report evaluation stages, where agents assess the quality of their own outputs and decide whether additional refinement is required. Together, these strategies enable DAR to combine structured orchestration with flexible, tool-augmented LLM reasoning.

The system implements memory tiers via ADK's session services: short-term memory maintains conversation context and query history within agent sessions; working memory shares state across agents through SessionService and stage variables.
\subsection{BigQuery Generative AI Integration}
The system leverages BigQuery's native AI functions: ML.GENERATE\_TEXT for natural language generation and data summarization, AI.GENERATE functionality for row-level text analysis and classification.
This in-database approach is employed because it: (1) eliminates data movement for AI operations, (2) leverages BigQuery's distributed computing infrastructure, (3) maintains data governance and security boundaries, and (4) reduces latency and cost compared to external API calls.

\section{Experimental setup}
This solution is implemented using Google ADK, an open-source framework for agentic programming. LLM agents are powered by Gemini \cite{gemini2024family, gemini2024onepointfive}. The code is available on Github\footnote{\url{https://github.com/MantisAnalytics/DAR}}.

To evaluate DAR, we designed a comparative case study in which a professional data analyst and a data scientist and the DAR system independently analyzed the same relational database under comparable conditions. The objective of this experiment was not to benchmark predictive accuracy, but to assess whether DAR can autonomously perform an end-to-end exploratory analysis and generate a written report within a practical time budget. The following subsections describe the dataset used in the study, the analysis task assigned to both agents, and the criteria employed to assess their performance.

\subsection{Dataset}

Our proof of concept utilized a relational database containing two interconnected tables. The dataset was selected to represent typical business intelligence scenarios requiring cross-table analysis, pattern identification, and insight generation.
The experimental dataset consists of two relational tables: an Assets table describing 26 critical facilities with 19 attributes, and an Incidents table containing 11,489 recorded events, also with 19 attributes. Table \ref{tab:asset_incident_dataset} summarizes the main attribute groups for each table, including identifiers, geographic information, descriptive fields, classification labels, and risk-related metadata that are used by DAR during autonomous exploration. The dataset is hosted on Kaggle\footnote{\url{www.kaggle.com/datasets/viktoriaskorik/incidents-dataset}}.

\begin{table}[t]
\centering
\scriptsize
\setlength{\tabcolsep}{3pt}  
\caption{Overview of the asset--incident dataset used in the experiments.}
\label{tab:asset_incident_dataset}
\begin{tabular}{lrrp{3.7cm}}
\toprule
Entity   & \# Records & \# Attr. & Attribute groups (examples) \\
\midrule
Assets   & 26         & 19       &
IDs (AssetID, OrganizationID, AssetName);\\
&&& security (ImpactRadius);\\
&&& geo (Latitude, Longitude, Country, City, Address);\\
&&& personnel (Headcount);\\
&&& contact (Phone, FocalPointName, Email, Phone);\\
&&& docs (DataSource, Hyperlink, Photo, ReportLink; 0 values = incomplete).\\[2pt]
Incidents & 11\,489   & 19       &
IDs (IncidentID, IncidentSourceId);\\
&&& temporal (IncidentDateTime, UTC);\\
&&& description (Title, IncidentDescription, Photo);\\
&&& geo (Latitude, Longitude, Country, City, Address, Region);\\
&&& classification (IncidentTypeName/ID, Relevance, eventCode);\\
&&& risk (SeverityLevel: 1=High, 2=Medium, 3=Low);\\
&&& docs (Hyperlink, DataSource).\\
\bottomrule
\end{tabular}
\end{table}
\subsection{Task Definition}
To ensure a fair comparison, both human analysts and the DAR system received an identical high-level assignment formulated as a short written brief. Specifically, they were instructed to perform exploratory data analysis on the provided dataset, identify significant patterns, trends, and anomalies, generate actionable insights from their findings, and produce a written report documenting the analysis and resulting recommendations. The instructions deliberately refrained from prescribing concrete queries, workflows, or modelling techniques, thereby allowing both human analysts and DAR to exercise their own strategies for exploration and sense-making within the same task constraints.

\section{Results}
Performance was assessed only based on the time required to complete the analysis and generate the final reports. We focus on time-to-insight as the primary criterion for three reasons. First, in practical decision-making contexts, the main bottleneck is often the latency between data availability and the delivery of usable findings; even modest reductions in turnaround time can have disproportionate impact on operational value. Second, DAR is intended as an enabler for rapid, first-pass exploration rather than a replacement for expert judgement, so its main promise lies in accelerating the initial analysis cycle rather than producing fundamentally different types of insights. Third, establishing a fine-grained ground truth for the “quality” or “correctness” of narrative analytical reports is inherently subjective and would require a separate large-scale human evaluation, which is beyond the scope of this study. For these reasons, we treat analysis-and-reporting time as the most informative and comparable metric for this initial evaluation. The results can be seen in Table \ref{tab:evaluation_results}

During the execution phase, the two conditions were run separately but under comparable constraints. A professional data analyst and a data scientist, with domain-relevant experience, conducted the exploration using conventional tools and methods of their choice (e.g., spreadsheet and business intelligence software), following the written assignment described above. In parallel, the DAR system was initialized with its default configuration and allowed to operate fully autonomously throughout the analysis phase, without manual intervention or prompt engineering beyond the initial task brief. Both human analysts and DAR worked on the same dataset and assignment, but executed their analyses independently to avoid mutual influence or information leakage.

The outputs produced by the human analyst and the DAR system differed primarily in depth and style rather than in overall focus. The human analyst delivered a thorough, highly detailed report that systematically reviewed the dataset, covered a broad range of analytical angles, and provided in-depth interpretations of the findings. Their report discussed subtle connections between variables, nuanced contextual explanations, and specific, expert-driven recommendations tailored to the use case.

In contrast, the DAR system generated a more concise report that emphasized breadth and clarity. Its output offered a structured summary of the key characteristics of the data, identified the main patterns and trends, and presented relevant insights supported by concrete evidence (e.g., descriptive statistics or query results). The recommendations formulated by DAR were practical but brief, focusing on the most salient findings rather than exhaustive coverage. Overall, human analysts’ output exhibited greater interpretive depth, whereas DAR prioritized succinct, high-level synthesis of the most important patterns and insights.

\begin{table}[t]
\centering
\caption{Evaluation results.}
\label{tab:evaluation_results}
\resizebox{\columnwidth}{!}{%
\begin{tabular}{lccc}
\toprule
\textbf{Metric} & \textbf{Human (AVG Time)} & \textbf{DAR} & \textbf{Ratio} \\
\midrule
Analysis time        & 5 hours 45 minutes    & 15 min & 23:1  \\
Report writing time  & 1 hour 25 minutes   & 1 min  & 85:1  \\
\midrule
\textbf{Total time}  & \textbf{7 hour 10 minutes} & \textbf{16 min} & \textbf{\(\sim 27{:}1\)} \\
\bottomrule
\end{tabular}%
}
\end{table}

\section{Discussion \& Conclusion}

We introduced the Data Agnostic Researcher (DAR), a multi-agent system for autonomous database exploration built entirely on in-database generative AI functions. We compared DAR with a data analysts and a data scientists on an asset-incident dataset under matched task instructions. DAR produced a coherent, insight-oriented report in a small fraction of the human analysis time, whereas the human output showed greater interpretive depth and contextual nuance.

Our experiments reveal a clear trade-off between speed and reliability. At its default configuration, DAR efficiently summarizes key patterns, trends, and anomalies. Extending runtime to push for deeper insights, however, increases errors and inconsistencies, including weakly supported findings and occasional contradictions across iterations. This suggests a practical “sweet spot”: DAR is most effective as a rapid, first-pass exploration tool rather than for prolonged, unconstrained analysis.

Overall, DAR completes the analytical task roughly $27\times$ faster than a human analyst (16 minutes vs.\ 7 hours 10 minutes), making it well suited for quick triage, initial data review, and screening multiple datasets. Humans remain better at uncovering subtle relationships and providing rich contextual interpretation. The most promising use of DAR is therefore as a front-end explorer that generates early hypotheses and highlights promising directions, which can then be refined and validated through targeted human analysis.
\section*{Limitations}
Our findings also indicate that DAR and human analysts are best viewed as complementary rather than competing resources. DAR excels at quickly scanning large relational datasets, surfacing prominent relationships, and producing an initial set of hypotheses and recommendations in an easily digestible format. Human analysts, in contrast, add value through comprehensive investigation, critical assessment of plausible but unverified patterns, and integration of domain knowledge and organizational context. In practical workflows, DAR can therefore serve as an autonomous “first reader” of the data, prioritizing areas of interest and reducing time-to-insight, while human experts focus on validating, extending, and contextualizing these preliminary findings for high-stakes decision-making. This division of labour aligns with a broader view of AI systems as accelerators of analytical sense-making rather than full replacements for human expertise.
\bibliography{custom}




\end{document}